\documentstyle[a4,epsfig,12pt]{article}

\newlength{\dinwidth}
\newlength{\dinmargin}
\newcommand{\ov}{\overline}
\begin{document}

\setcounter{topnumber}{1}
\setcounter{bottomnumber}{1}
\title{
\vskip-1cm{\baselineskip16pt
\centerline{\normalsize DESY 94-035\hfill ISSN 0418-9833}
\centerline{\normalsize WIS-94/15/MAR-PH \hfill}
\centerline{\normalsize March 1994\hfill}}
\vskip3cm
Production and Decay of the Standard Model Higgs Boson at LEP200}
\date{}

\author{Eilam Gross,$\!\mathstrut^{\dag}$
        Bernd A. Kniehl,$\!\mathstrut^{\ddag}$
        and Gustavo Wolf$\mathstrut^{\dag}$\\
{\small $\dag$ Department of Particle Physics}\\
{\small Weizmann Institute of Science, Rehovot 76100, Israel}\\
{\small $\ddag$ II.~Institut f\"ur Theoretische Physik,
 Universit\"at Hamburg}\\
{\small Luruper Chaussee 149, 22761 Hamburg, Germany}}
\maketitle

\begin{abstract}
We collect and update theoretical predictions for the production rate
and decay branching fractions of the Standard Model Higgs boson that will
be relevant for the Higgs search at LEP200.
We make full use of the present knowledge of radiative corrections.
We estimate the systematics arising from theoretical and experimental
uncertainties.
\bigskip
\bigskip
\bigskip
\end{abstract}

  \clearpage
\section{Introduction}

     The main mechanism by which the Higgs boson is produced in
 $e^+e^-$ collisions at LEP is the Higgsstrahlung process where
 the primary $Z$ boson, {\it i.e.}, the one that is
 produced by $e^+e^-$ annihilation,
 radiates a Higgs boson and then decays into a fermion
 pair~\cite{Bjorken,Ellis}.

    At LEPI, the primary $Z$ boson is on resonance,
  $e^+e^-\to Z\to HZ^*\to Hf\bar f$ (Bjorken process);
  see Fig.~\ref{FIG-FEYN1}a.
\begin{figure}[h]
\caption{Feynman diagrams for Higgs-boson production at (a) LEPI and
(b) LEP200.}
\label{FIG-FEYN1}
\end{figure}
 As a consequence, the production cross section decreases dramatically
 with $m_H$ increasing, and the search potential of LEPI
 approaches its sensitivity limit, which is probably at $m_H=65$~GeV or
 so~\cite{Gross-Yepes}.

 At LEP200, operating at centre-of-mass (CM) energies in the range
 $\sqrt s=170$--200~GeV, the
 primary $Z$ boson is virtual, while the secondary $Z$ boson, {\it i.e.},
 the one that coexists with the Higgs boson, is close to its mass shell,
 $e^+e^-\to Z^*\to HZ\to Hf\bar f$; see Fig.~\ref{FIG-FEYN1}b.
  As a result, the production cross section is enhanced when $\sqrt s$
 is large enough to allow both the Higgs and $Z$ bosons to be
 on their mass shells. LEP200 is, therefore, the suitable
 machine to look for Higgs bosons, up to about
 $m_H=80$--100~GeV, depending on the available CM
 energy~\cite{LEP200}.

A dedicated study of the Higgs-boson production cross section at LEP200
does not exist in literature.
The
 Born cross section of $e^+e^-\to Hf\bar f$ was calculated analytically
by Berends and Kleiss~\cite{Berends-Kleiss}. First-order radiative
corrections appropriate to LEPI energies were then
calculated using the so-called Improved Born Approximation
(IBA)~\cite{IBA} and complemented by a
specific heavy-top-quark
correction to the $ZZH$ vertex~\cite{Hioki}.
The full one-loop radiative corrections to the four-point process
$Z\to Hf\bar f$ were
 calculated in Ref.~\cite{Kniehl-LEP1}, where it was shown
that, for LEPI energies, the IBA along with the $ZZH$ correction
agrees with the fully
 corrected result for $\Gamma\left(Z\to Hf\bar f\right)$
at the level of 1\%.
  However for LEP200 energies, this calculation is
  no longer adequate because the primary $Z$-boson
  is virtual. In principle, one would like
  to know the full radiative corrections to the five-point process
  $e^+e^-\to Hf\bar f$, but this happens to be very cumbersome and
  has not been tackled yet.
  For $\sqrt s>m_H+m_Z$, one may resort to the four-point process
  $e^+e^-\to HZ$, for which the one-loop radiative corrections are
  known~\cite{Kniehl-LEP2,FJ}.
  However, this calculation does not take into account
  finite-width effects of the secondary $Z$ boson,
  which is indeed close to its mass shell,
  but, nevertheless, has an observable width.

In Sect.~2, we make an attempt to
predict the five-point process as accurately as possible.
To this end, we incorporate in the tree-level result the
effects of initial-state electromagnetic bremsstrahlung
to second order, including exponentiation of the infrared-sensitive
parts.
We then plug in the
  known weak corrections to the four-point process
along with experimental information on the $Z$-boson width.
In the limit of $M_t\gg \sqrt s$, the weak corrections are dominated
by virtual top-quark contributions.
The IBA naturally accounts for most of these terms, except for the one
that
 arises
  from the one-loop renormalization of the $HZZ$ vertex~\cite{Hioki}.
The latter must be added by hand.
We compare our best estimate with the IBA-type evaluation and find
that the two differ appreciably for $\sqrt s\gg m_H+m_Z$.
Incidentally, the two approaches agree quite well close to threshold.
The IBA is applicable also below threshold. On the other hand, the cross
section drops rapidly below threshold, so that only $\sqrt s$ values
down to
 a few times $\Gamma_Z$ below $m_H+m_Z$ are relevant phenomenologically.
We thus argue that one may still conservatively
use the threshold value of the full weak
correction to the four-point process in that range below threshold,
and estimate the theoretical uncertainty of the result obtained.

   In Sect.~3, the Higgs-boson decay branching
   ratios are re-evaluated and compared with previous results.
     Many experimental
     Higgs-search papers quote the Higgs-boson branching ratio
    to $b\bar b$ as 85-87\% and that
    to $\tau^+\tau^-$ as  5-8\%~\cite{LEP200,brs}.
    Both branching fractions are of extreme experimental importance.
  In particular at LEP200, the success of the search for the Higgs boson
  relies on the $b$-tagging capability of the experiment. It is obvious
  that a reliable prediction of $BR\left(H\rightarrow b\bar b\right)$
  is crucial.

 The Higgs bosons accessible at LEP200 ($m_H<100$~GeV) are relatively
 long-lived, with $\Gamma_H$ of the order of a few MeV, and can thus
 be taken to be on mass shell in the analysis.
 The experimentally relevant quantities are thus the
 total production rate and
 the various branching fractions including their radiative corrections.
 Most of the theoretical papers on radiative
 corrections to Higgs-boson decays
 consider partial decay widths rather than branching ratios; for a recent
 review, see Ref.~\cite{Kniehl}.
 One purpose of the present paper is to update the Higgs branching ratios
 relevant for LEP200 making full use of the present knowledge of
 radiative corrections.
 The calculations will be shown here in detail.

 \section{Higgs-Boson Production Cross Section}
 In this section, we evaluate the radiatively corrected cross section
 of $e^+e^-\to Hf\bar f$
  using both the IBA and the full weak corrections to
 $\sigma(e^+e^-\to HZ)$.
 Before doing so, we shall briefly review the Born cross section and
 initial-state bremsstrahlung, which are the core of
 the analysis.

 \subsection{Born Approximation}

 We choose to work in the modified on-mass-shell (MOMS) scheme, in
 which the Born amplitude is expressed in terms of the Fermi
 constant, $G_F$, measured
  in muon decay so as to suppress large logarithms
 due to virtual light charged fermions in the radiative corrections.
 The weak mixing angle is defined as $\sin^2\theta_W=1-m_W^2/m_Z^2$.
 For given $m_H$ and $M_t$,
  $m_W$ is fine-tuned such that the perturbative
 calculation of the muon lifetime agrees with its high-precision
 measurement, {\it i.e.}, such that
 \begin{equation}
  \label{GFermi}
  G_F=\frac{\pi\alpha}{\sqrt 2\sin^2\theta_Wm_W^2}\,
  \frac{1}{1-\Delta r(m_W,m_H,M_t)}
  \end{equation}
  is satisfied.
  Here $\Delta r$~\cite{Delta-r}
   embodies those radiative corrections to the
  muon decay width that the Standard Model generates on top of the
  traditional calculation within the QED-improved Fermi Model.
  For consistency of our analysis, we use $\Delta r$ in the one-loop
  approximation~\cite{Delta-r}.
  For a discussion of $\Delta r$ beyond one loop, we refer to
  Ref.~\cite{Halzen}.

 In the MOMS scheme, the Born
  cross section of $e^+e^-\to Hf\bar f$, where
 all fermion flavours $f\ne t$ are summed over, may be written
 as~\cite{Berends-Kleiss}
 \begin{equation}
 \label{EQ-BORN}
 \sigma_{\mbox{\protect\scriptsize Born}}^0(s)=
 \frac{G_F^2 (v_e^2+1)m_Z^3\Gamma_Z}{96\pi^2 s}\,
 \frac{m_Z^2/s}{(1-m_Z^2/s)^2+b^2}
 \int_{x_1}^{x_2}dx\,{\cal F}(x),
 \end{equation}
 where
 \begin{equation}
 {\cal F}(x)=\frac{(12+2a-12x+x^2)\sqrt{x^2-a}}
 {(x-x_p)^2+b^2},
\end{equation}
$v_e=4\sin^2\theta_W-1$,
$a=4m_H^2/s$, $b=m_Z\Gamma_Z/s$, $x=2E_H/\sqrt s$,
$x_p=1+(m_H^2-m_Z^2)/s$, $x_1=\sqrt a$, $x_2=1+\frac{1}{4}a$,
and $E_H$ is the Higgs-boson energy in the laboratory frame.
Here the $Z$-boson propagators are written in the Breit-Wigner form
with $m_Z=(91.187\pm 0.007)$~GeV and
$\Gamma_Z=(2.489\pm 0.007)$~GeV~\cite{LEP-WEAK}.
The integral is performed analytically by means of complex
analysis~\cite{Berends-Kleiss}.

 \subsection{Bremsstrahlung}

 Initial-state corrections to $e^+e^-$ annihilation are of prime
 importance. These are available to $O(\alpha^2)$ in
 QED~\cite{ISR1,ISR2}. Here we adopt the formalism of Ref.~\cite{ISR2}
 taking into account real and virtual contributions due to photons and
 additional $e^+e^-$ pairs.
 This is achieved
  by convoluting $\sigma_{\mbox{\protect\scriptsize Born}}^0$
 with the appropriate radiator function over the full range of
 center-of-mass
  energies, $\sqrt{s^\prime}$, accessible after bremsstrahlung.
 Specifically,
 \begin{equation}
 \label{expo}
 \sigma_{\mbox{\protect\scriptsize Born}}(s)=\int_{x_0}^1 dx\,G(x)
 \sigma_{\mbox{\protect\scriptsize Born}}^0(xs),
 \end{equation}
 where $x=s^\prime/s$ and $x_0=m_H^2/s$, taking final-states fermions to
 be massless.
 The resummation of infrared-sensitive contributions is accomplished
 by writing the radiator function, $G(x)$, in an exponentiated
 form~\cite{ISR2},
 \begin{equation}
 G(x)=\beta (1-x)^{\beta-1}\delta^{V+S}+\delta^H(x),
 \end{equation}
 where $\delta^{V+S}$ and $\delta^H(x)$ are polynomials in
 $L=\ln\left(s/m_e^2\right)$ and $\beta=(2\alpha/\pi)(L-1)$.
 Note that here
  $\alpha =1/137.035...$ because the emitted photons are real.
 The term $\delta^{V+S}$ collects virtual and soft contributions,
 while $\delta^H$ originates from hard bremsstrahlung.
 For further details, see Ref.~\cite{ISR2}.

 In Fig.~\ref{FIG-3},
  $\sigma_{\mbox{\protect\scriptsize Born}}$ is plotted
 versus
 $\sqrt s$ for $m_H=80$~GeV (dotted curve).
 Numerical results for selected values of $\sqrt s$ and $m_H$ are listed
 in the third column of Table~\ref{TAB-1}.
\begin{figure}
\caption{
$\sigma_{\mbox{\protect\scriptsize Born}}$ (dotted line),
$\sigma_{\mbox{\protect\scriptsize IBA}}$ (dashed line), and
$\sigma_{\mbox{\protect\scriptsize full}}$ (solid line)
versus $\protect\sqrt s$ for $m_H=80$~GeV assuming $M_t=165$~GeV.}
\label{FIG-3}
\end{figure}
 \begin{table}[tbp]
  \begin{center}
    \begin{tabular}{|c||c||c|c|c||}\hline
 $\sqrt s$ [GeV] &
  $m_H$ [GeV] & $\sigma_{\mbox{\protect\scriptsize Born}}$ [pb]
 &
 $\sigma_{\mbox{\protect\scriptsize IBA}}$ [pb] &
 $\sigma_{\mbox{\protect\scriptsize full}}$
 [pb]\\ \hline \hline
   170 & 60 & 1.171    & 1.158    & 1.139    \\ \hline
       & 70 & 0.708    & 0.700    & 0.692    \\ \hline
       & 80 & 0.076    & 0.075    & 0.075    \\ \hline
       & 90 & 0.011    & 0.011    & 0.011    \\ \hline
       &100 & 0.004    & 0.004    & 0.004    \\ \hline
     \hline
   180 & 60 & 1.119    & 1.106    & 1.083    \\ \hline
       & 70 & 0.834    & 0.825    & 0.811    \\ \hline
       & 80 & 0.501    & 0.496    & 0.490    \\ \hline
       & 90 & 0.053    & 0.053    & 0.052    \\ \hline
       &100 & 0.008    & 0.008    & 0.008    \\ \hline
    \hline
   190 & 60 & 1.016    & 1.004    & 0.980   \\ \hline
       & 70 & 0.821    & 0.812    & 0.795    \\ \hline
       & 80 & 0.610    & 0.603    & 0.593    \\ \hline
       & 90 & 0.365    & 0.361    & 0.357    \\ \hline
       &100 & 0.038    & 0.038    & 0.038    \\ \hline
    \hline
   200 & 60 & 0.905    & 0.895    & 0.871    \\ \hline
       & 70 & 0.764    & 0.756    & 0.738    \\ \hline
       & 80 & 0.616    & 0.609    & 0.597    \\ \hline
       & 90 & 0.456    & 0.451    & 0.443    \\ \hline
       &100 & 0.272    & 0.269    & 0.265    \\ \hline
    \hline
   210 & 60 & 0.802    & 0.793    & 0.770    \\ \hline
       & 70 & 0.696    & 0.688    & 0.670    \\ \hline
       & 80 & 0.586    & 0.579    & 0.566    \\ \hline
       & 90 & 0.471    & 0.465    & 0.456    \\ \hline
       &100 & 0.347    & 0.343    & 0.337    \\ \hline
    \hline
          \end{tabular}\newline
  \end{center}
  \caption[Table 1]
  {\label{TAB-1}
$\sigma_{\mbox{\protect\scriptsize Born}}$,
$\sigma_{\mbox{\protect\scriptsize IBA}}$, and
$\sigma_{\mbox{\protect\scriptsize full}}$
for several values of $\sqrt s$ and $m_H$~GeV assuming $M_t=165$~GeV.}
\end{table}

\subsection{Improved Born Approximation}

In order to take into account the running of $\alpha$ and virtual
heavy-top-quark effects in the evaluation of the Higgs-boson production
cross section, it became a common practice among the LEP collaborations
to use the
 IBA~\cite{IBA} complemented by the specific $ZZH$ vertex correction
of top origin~\cite{Hioki}.
In the IBA,
 the $Z$-boson and photon propagators get dressed with self-energy
insertions,
 which comprise the leading effects due to light charged fermions
and a heavy top quark.
These effects may be accommodated by introducing effective parameters,
{\it i.e.}, by substituting
$\alpha\to \ov{\alpha}=\alpha(m_Z)$ and
$\sin^2\theta_W \to\sin^2\ov\theta_W=1-\cos^2\ov\theta_W=
\sin^2\theta_W+\cos^2\theta_W\Delta\rho$
in the Born approximation of the on-mass-shell scheme, where
\begin{equation}
1-\frac{1}{\rho}=\Delta\rho =\frac{3G_FM_t^2}{8\pi^2\sqrt 2}.
\end{equation}
Using the relation
\begin{equation}
   \rho G_F=\frac{\pi\ov{\alpha}}
  {\sqrt 2\sin^2\ov\theta_W\cos^2\ov\theta_Wm_Z^2},
\end{equation}
which
 emerges from Eq.~(\ref{GFermi}) by retaining only the dominant terms
of $\Delta r$,
the IBA cross
 section is obtained from Eq.~(\ref{EQ-BORN}) by substituting
$G_F\to\rho G_F\approx(1+\Delta\rho)G_F$
and $\sin^2\theta_W\to\sin^2\ov\theta_W$
and including
 the overall factor $(1-\frac{8}{3}\Delta\rho)$~\cite{Hioki} to
account for the $ZZH$ vertex correction.
The result reads
 \begin{equation}
 \sigma_{\mbox{\protect\scriptsize IBA}}^0(s)=
 \frac{G_F^2(\ov v_e^2+1)m_Z^3\Gamma_Z}{96\pi^2 s}\,
 \frac{m_Z^2/s}{(1-m_Z^2/s)^2+b^2}
 (1+2\Delta\rho)(1-\frac{8}{3}\Delta\rho)
 \int_{x_1}^{x_2}dx\,{\cal F}(x),
 \end{equation}
 where $\ov v_e=4\sin^2\ov\theta_W-1$.
Convolution
 of $\sigma_{\mbox{\protect\scriptsize IBA}}^0$ with $G$ according to
Eq.~(\ref{expo}) leads to $\sigma_{\mbox{\protect\scriptsize IBA}}$,
which is potted in Fig.~\ref{FIG-3}, too (dashed curve).
Numerical results may be found in the fourth column of Table~\ref{TAB-1}.

\subsection{Full One-Loop Radiative Corrections}

 In the case of $Z\to Hf\bar f$ relevant for LEPI, the IBA is in
 reasonable
  agreement with the full one-loop calculation~\cite{Kniehl-LEP1},
 especially in the upper $M_t$ range.
 This may be understood by observing that, close to $Z$-boson peak,
 the most significant contributions arise from loop amplitudes with
 resonant
  propagators, the leading terms of which are retained in the IBA.
 However, this is not necessarily the case at LEP200 energies.
 In fact, it has been shown~\cite{Kniehl-LEP2,FJ} that
 $\sigma(e^+e^-\to HZ)$
  receives sizeable contributions from box diagrams,
 which do not enter the IBA.

 One should keep in mind that in Refs.~\cite{Kniehl-LEP2,FJ}
 the $Z$ boson was treated as a stable particle, so that there is
 no cross section for $\sqrt s<m_Z+m_H$.
 For $\sqrt s$ values a few energies above $m_H+m_Z$, this treatment is
 expected to be a fair one.
 However, at LEP200, the Higgs detection sensitivity might be pushed
 to phase space regions where the number of expected Higgs events
 is very modest, and special care must be
  exercised.
    This can be seen in Fig.~\ref{FIG-3A}, where the probability that
 the secondary $Z$ boson is produced with mass $m_{f\bar f}$ is shown for
 $\sqrt s=180$~GeV.
 This probability
  is obtained by normalizing the differential cross section,
 \begin{equation}
 \label{masspec}
 \frac{d\sigma_{\mbox{\protect\scriptsize Born}}^0}{d m_{f\bar f}^2}
\propto
\frac{p_{f\bar f}\left(3m_{f\bar f}^2+p_{f\bar f}^2\right)}
{\left[\left(s-m_Z^2\right)^2+m_Z^2\Gamma_Z^2\right]
\left[\left(m_{f\bar f}^2-m_Z^2\right)^2+m_Z^2\Gamma_Z^2\right]},
\end{equation}
where
$p_{f\bar f}
=\left(\lambda\left(s,m_{f\bar f}^2,m_H^2\right)/4s\right)^{1/2}$,
with $\lambda(a,b,c)=a^2+b^2+c^2-2(ab+bc+ca)$,
is the $f\bar f$ three-momentum in the CM frame.
\begin{figure}
\caption{Distributions
 of the probability that the $f\bar f$ pair produced
through $e^+e^-\to Hf\bar f$
 at $\protect\sqrt s=180$~GeV has invariant mass
$m_{f\bar f}$ for $m_H=70$, 80, and 90~GeV.}
\label{FIG-3A}
\end{figure}
    One can clearly see that there is a finite, yet small probability
    for the $Z$-boson to be off-shell, especially as $m_H$ approaches
    the threshold value $\sqrt s-m_Z$.

   In the following, we suggest a way to incorporate finite-width effects
   in the one-loop
    calculation of the Higgs-boson production cross section
   at LEP200.
      The calculation of
     the full radiative corrections to the $2\to3$ process
     $e^+e^-\to Hf\bar f$ requires the computation of an enormous number
     of Feynman diagrams and does not exist in literature.
However, at least
for $\sqrt s>m_H+m_Z$, one can take advantage of the fact
that the secondary $Z$-boson is preferably on-shell, so that the
radiative corrections factorize approximately into a part connected
with the $2\to2$ process $e^+e^-\to HZ$ and one related to the
subsequent $Z$-boson decay, which are both known.
The second part may be included elegantly by using the experimental
value of $\Gamma_Z$, which is radiatively corrected by nature.
Taking into account also initial-state bremsstrahlung, we may write,
for $\sqrt s>m_H+m_Z$,
     \begin{eqnarray}
   \label{EQ-FULL1}
  \sigma_{\mbox{\protect\scriptsize full}}(s)
  &=&\left(1+2{\cal R}e\Delta_{\mbox{\protect\scriptsize weak}}(s)\right)
  \int_{x_0}^1dx\,G(x)\sigma_{\mbox{\protect\scriptsize
   Born}}^0(xs)\nonumber\\
  &=&\left(1+2{\cal R}e\Delta_{\mbox{\protect\scriptsize weak}}(s)\right)
  \sigma_{\mbox{\protect\scriptsize Born}}(s),
   \end{eqnarray}
   where $\Delta_{\mbox{\protect\scriptsize weak}}$ is the finite,
   gauge-invariant weak correction to $\sigma(e^+e^-\to HZ)$ as given by
   Eq.~(4.8) of Ref.~\cite{Kniehl-LEP2}.
   One might wonder whether the weak correction term should be
   evaluated at
    $xs$ and included as part of the integrand. We have chosen
   not to do so.
   However, this question becomes irrelevant by noticing that this
   modification changes the cross section by less than 0.1\%.

   For $\sqrt s<m_H+m_Z$, one faces the problem that
   $\Delta_{\mbox{\protect\scriptsize weak}}$ is not defined, since the
 secondary
   $Z$ boson is pushed from its mass shell.
   For the sake of continuity, we propose to use Eq.~(\ref{EQ-FULL1})
   with
   $\Delta_{\mbox{\protect\scriptsize weak}}$ evaluated at $(m_H+m_Z)^2$
   instead of $s$.
   This
   procedure turns out to be a conservative one as will be shown below.

   In Fig.~\ref{FIG-3},
    $\sigma_{\mbox{\protect\scriptsize full}}$ (solid line)
 is
   compared with $\sigma_{\mbox{\protect\scriptsize Born}}$ and
   $\sigma_{\mbox{\protect\scriptsize IBA}}$.
   Numerical values are listed in the last column of Table~\ref{TAB-1}.
   We observe that the IBA overshoots the full one-loop calculation
   for $\sqrt s\gg m_H+m_Z$.
   To elaborate this point, we show in Fig.~\ref{FIG-4} the ratio
   $\sigma_{\mbox
   {\protect\scriptsize full}}/\sigma_{\mbox{\protect\scriptsize
 IBA}}$
   as a function of $\sqrt s$.
\begin{figure}
\caption{$\sigma_{\mbox{\protect\scriptsize
 full}}/\sigma_{\mbox{\protect\scriptsize IBA}}$
versus $\protect\sqrt s$ for $m_H=80$~GeV assuming $M_t=165$~GeV.}
\label{FIG-4}
\end{figure}
 It is clearly seen
 that the IBA rapidly deteriorates as $\sqrt s$ increases.
 On the other hand, the ratio gets close to one as $\sqrt s$ approaches
 $m_H+m_Z$, the value at threshold being 0.995.
 For $\sqrt s<m_H+m_Z$, the ratio is independent of $\sqrt s$, since
 we continue to use
 the threshold value of $\Delta_{\mbox{\protect\scriptsize
 weak}}$
 in the evaluation of $\sigma_{\mbox{\protect\scriptsize full}}$.
 The corrections
 implemented in the IBA are independent of $\sqrt s$ anyway.
 Thus,
  $\sigma_{\mbox{\protect\scriptsize IBA}}$ is perfectly well defined
 theoretically also below threshold.
 Nevertheless,
  for the sake of a continuous description, we are in favour of
 using
  $\sigma_{\mbox{\protect\scriptsize full}}$, provided that $\sqrt s$
 is not more than a few times $\Gamma_Z$ below $m_H+m_Z$.
 This attitude is also conservative from the experimental point of view,
 since $\sigma_{\mbox{\protect\scriptsize IBA}}$ might overestimate the
 true production rate. However, since
  both approaches agree to the level of 0.5\%, we do not anticipate
 a great theoretical uncertainty.
 However, a firm conclusion concerning the virtue of this approximation
 can be drawn only from a full one-loop calculation of
 $\sigma\left(e^+e^-\to Hf\bar f\right)$.
In Fig.~\ref{FIG-3}, we assumed $m_H=80$~GeV.
In Fig.~\ref{FIG-5}, we show the $\sqrt s$ dependence of
$\sigma_{\mbox{\protect\scriptsize full}}$
also for other values of $m_H$.
\begin{figure}
\caption{$\sigma_{\mbox{\protect\scriptsize full}}$
versus $\protect\sqrt s$ for $m_H=70$, 80, 90, and 100~GeV assuming
$M_t=165$~GeV.}
\label{FIG-5}
\end{figure}
In practice,
 $\sqrt s$ will be fixed at some value between 170 and 200~GeV.
It is then useful to know the $m_H$ dependence of
$\sigma_{\mbox{\protect\scriptsize full}}$,
which is shown in Fig.~\ref{FIG-6}.
\begin{figure}
\caption{$\sigma_{\mbox{\protect\scriptsize full}}$
versus $m_H$ for $\protect\sqrt s=170$, 180, 190, and 200~GeV assuming
$M_t=165$~GeV.}
\label{FIG-6}
\end{figure}
\subsection{Systematics}
The accuracy of the predicted value of
$\sigma_{\mbox{\protect\scriptsize full}}$
is primarily limited by the errors on the input parameters.
Apart from $m_H$, these are $m_Z$, $\Gamma_Z$, and $M_t$.
$m_Z$ and $\Gamma_Z$ have been measured at LEP to high accuracy,
$m_Z=(91.187\pm0.007)$~GeV
and $\Gamma_Z=(2.489\pm 0.007)$~GeV~\cite{LEP-WEAK}.
Recent global analyses~\cite{LEP-WEAK} of LEP data suggest
$M_t=(166^{+17+19}_{-19-22})$~GeV
via loop effects. Recently, the D0 Collaboration at the Fermilab Tevatron
has announced a lower limit of 131~GeV on $M_t$~\cite{D0}.
Therefore, $M_t=(165\pm 35)$~GeV covers the most probable $M_t$ range.

Changing $m_Z$ ($\Gamma_Z$) by $\pm  2\sigma$ shifts
$\sigma_{\mbox{\protect\scriptsize full}}$
by at most 0.6\% (0.5\%);
the maximum effect occurs when $\sqrt s=m_H+m_Z$ ($\sqrt s<m_H+m_Z$).
The variation of $\sigma_{\mbox{\protect\scriptsize full}}$ with $M_t$ is
studied in Fig.~\ref{FIG-7} for typical LEP200 conditions,
$\sqrt s=180$~GeV and $m_H=70$~GeV.
We see that
 the present uncertainty in $M_t$ induces a systematic error of
$\pm0.5\%$ in $\sigma_{\mbox{\protect\scriptsize full}}$.
\begin{figure}
\caption{
$\sigma_{\mbox{\protect\scriptsize full}}$ versus $M_t$
for $\protect\sqrt s=180$~GeV and $m_H=70$~GeV.}
\label{FIG-7}
\end{figure}
Since the one-loop corrections to the Higgs-boson production rate
at LEP200 are
 relatively modest in the MOMS scheme, below 4\% in magnitude,
we expect that
 the theoretical uncertainty due to unknown higher orders is
insignificant.
In summary, we estimate the total systematic error on the Higgs-boson
production cross
 section to be of the order of 0.9\%.

 \section{Higgs-Boson Decay Branching Ratios}

In this
 section, we study in detail the branching ratios of the Higgs boson
that are relevant at LEPI and LEP200.
We start by reviewing the tree-level results.
We then analyze the influence of radiative corrections.
Finally, we estimate the systematic errors involved in the calculations.

 \subsection{Born Approximation}

The Higgs
 boson couples directly to fermions thereby generating their masses.
At tree level, the coupling strength is $2^{1/4}G_F^{1/2}m_f$ and the
$H\to f\bar f$ decay width is
  \begin{equation}
  \label{EQ-Gamma}
  \Gamma_0\left(H\rightarrow f\bar f\right)=
  \frac{N_cG_Fm_Hm_f^2\beta_f^3}{4\pi\sqrt2},
  \end{equation}
 where $N_c=1$ (3) for leptons (quarks) and
 $\beta_f=\sqrt{1-4m_f^2/m_H^2}$ is the velocity of $f$ in the CM frame.
 Thus, the tree-level branching fractions are
 \begin{equation}
  \label{EQ-BRs}
 BR\left(H\rightarrow f\bar f\right) = \frac
 {N_c m_f^2 \beta_f^3}
 {\sum_{f'\ne t} N_c m_{f'}^2 \beta_{f'}^3}.
 \end{equation}
 Using
  $m_b=4.7$~GeV, $m_c=1.45$~GeV~\cite{Q-masses}, and $m_\tau=1.777$~GeV,
 one finds\break $BR\left(H\rightarrow b\bar b\right)=87\%$,
 $BR\left(H\rightarrow c\bar c\right)=9\%$, and
 $BR\left(H\rightarrow\tau^+\tau^-\right)=4\%$.
These
lowest-order estimates are subject to electroweak and QCD corrections.
Moreover, higher-order decay channels need to be taken into account.

 \subsection{Higgs-Boson Decays to Two Electroweak Bosons}
We start by considering the decay of the Higgs boson to four fermions via
a pair of virtual $W$ bosons.
This channel becomes relevant for LEP200 as soon as $m_H>m_W$, so that
one $W$ boson can get on its mass shell.
Its partial decay width can be written as~\cite{Kniehl90}
\begin{eqnarray}
\label{hwwrate}
\Gamma(H\to W^*W^*)&=&
\int_{\sqrt {s_+}+\sqrt{s_-}\leq m_H} \frac{ds_+ds_-}{\pi^2}\,
\frac{s_+\Gamma_W/m_W}
{\left( s_+-m_W^2 \right)^2+m_W^2\Gamma_W^2} \nonumber \\
&&\times{}
\frac{s_-\Gamma_W/m_W}
{\left( s_--m_W^2 \right)^2+m_W^2\Gamma_W^2}
\Gamma\left(m_H^2,s_+,s_-\right),
\end{eqnarray}
where
\begin{equation}
\Gamma\left(m_H^2,s_+,s_-\right)=\frac{3G_Fm_W^4}{2\pi\sqrt2m_H^3}
\sqrt{\lambda\left(m_H^2,s_+s_-\right)}
\left(1+\frac{\lambda\left(m_H^2,s_+s_-\right)}{12s_+s_-}\right),
\end{equation}
where $\lambda(a,b,c)$ is defined below Eq.~(\ref{masspec})
and we use the experimental values
$\Gamma_W=2.12$~GeV and $m_W=80.22$~GeV~\cite{PDG-alphaS}.
The formula for the $H\to Z^*Z^*$ decay width emerges from
Eq.~(\ref{hwwrate})
by substituting
 $m_W$ and $\Gamma_W$ by $m_Z$ and $\Gamma_Z$, respectively,
and including
the factor 1/2 to account for identical-particle symmetrization.

The $H\to\gamma\gamma$~\cite{Ellis,Vainshtein} and
$H\to\gamma Z$~\cite{Barroso} decays proceed through $W$-boson and
charged-fermion loops and are generally less significant for the Higgs
search at LEP 200.
QCD corrections to their partial widths are well under
control~\cite{gammagamma,gammaZ}.

 \subsection{QCD and Electroweak Corrections}

 In the case of
  $H\to q\bar q$, it is important to include QCD corrections
 to Eq.~(\ref{EQ-Gamma}) \cite{Braaten,Gorishny}.
 In fact, when the pole mass, $M_q$, is used as a basic parameter,
 these corrections contain large logarithms of the form
 $(\alpha_s/\pi)^n\ln^m\left(m_H^2/M_q^2\right)$, with $n\ge m$.
 Appealing to the renormalization-group equation, these logarithms may be
 absorbed completely into the running quark mass, $m_q(\mu)$, evaluated
 at scale $\mu=m_H$.
 In this way, these logarithms are resummed to all orders and the
 perturbation expansion converges more rapidly.
 This observation gives support to the notion that the $Hq\bar q$ Yukawa
 couplings are controlled by the running quark masses.

 The values of $M_q$ may be estimated from QCD sum rules.
 In our analysis, we use $M_c=(1.45\pm0.05)$~GeV and
 $M_b=(4.7\pm0.2)$~GeV~\cite{Q-masses}.
 To obtain $m_q(m_H)$, we proceed in two steps.
 Firstly, we evaluate $m_q(M_q)$ from
 \begin{equation}
   m_q(M_q)=\frac{M_q}{1+(4/3)\alpha_S(M_q)/\pi+
   K(\alpha_S(M_q)/\pi)^2},
 \end{equation}
 with~\cite{Gray}
 \begin{equation}
  K\approx~16.11-1.04\sum_{i=1}^{n_F-1}\left(1-\frac{M_i}{M}\right),
 \end{equation}
 where the sum extents over all quark flavours with $M_i<M_q$.
 Specifically, $K_c=13.2$ and $K_b=12.3$.
 Secondly, we determine $m_q(m_H)$ via the scaling law
 \begin{equation}
 m_q(m_H)=m_q(M_q)\frac{c_q(\alpha_S(m_H)/\pi)}{c_q(\alpha_S(M_q)/\pi)},
 \end{equation}
 where
 \begin{eqnarray}
 c_c(x)&=&\left(\frac{25}{6}x\right)^{12/25}(1+1.014x+1.389x^2),\\
 c_b(x)&=&\left(\frac{23}{6}x\right)^{12/23}(1+1.175x+1.501x^2).
 \end{eqnarray}

 Throughout our analysis, we evaluate $\alpha_S$ from the two-loop
 formula~\cite{PDG-alphaS},
 \begin{equation}
 \alpha_S(\mu)=\frac{12\pi}
 {(33-2n_F)\ln\left(\mu^2/\Lambda_{(n_F)}^2\right)}
 \left[1-\frac{6(153-19n_F)}{(33-2n_F)^2}\,
 \frac{\ln\left[\ln\left(\mu^2/\Lambda_{(n_F)}^2\right)\right]}
 {\ln\left(\mu^2/\Lambda_{(n_F)}^2\right)}\right],
 \end{equation}
 where $n_F$ is the number of quark flavours active at scale $\mu$
 and $\Lambda_{(n_F)}$ is the appropriate asymptotic scale parameter.
 We fix $\Lambda_{(5)}$
  by requiring that $\alpha_S(m_Z)=0.123$~\cite{AlphaS}
 and determine $\Lambda_{(4)}$ from the condition that $\alpha_s(\mu)$ be
 continuous at the flavour threshold $\mu=M_b$.
 Using the above value of $M_b$, we find $\Lambda_{(4)}=0.416$~GeV
 and $\Lambda_{(5)}=0.296$~GeV.
 The scale dependences of $m_c$ and $m_b$ to $O\left(\alpha_S^2\right)$
 are illustrated in Fig.~\ref{FIG-1} and Table~\ref{TAB-11}.
\begin{figure}
\caption{$\mu$ dependence of (a) $m_c(\mu)$ and (b) $m_b(\mu)$ evaluated
to $O\left(\alpha_S^2\right)$ assuming $M_c=1.45$~GeV, $M_b=4.7$~GeV, and
$\alpha_S(m_Z)=0.123$.}
\label{FIG-1}
\end{figure}
\begin{table}[tbp]
  \begin{center}
    \begin{tabular}{|c|c|c||}\hline
    $m_H$ [GeV] & $m_b(m_H)$ [GeV] & $m_c(m_H)$ [GeV] \\ \hline \hline
     50            &   2.89         &   0.51         \\ \hline
     60            &   2.84         &   0.50         \\ \hline
     70            &   2.80         &   0.50         \\ \hline
     80            &   2.77         &   0.49         \\ \hline
     90            &   2.74         &   0.48         \\ \hline
    100            &   2.71         &   0.48         \\ \hline
    \hline\end{tabular}\newline
  \end{center}
  \caption[The Running Quark Masses]
  {\label{TAB-11}
$m_b(m_H)$ and $m_c(m_H)$ evaluated to $O\left(\alpha_S^2\right)$
for several values of $m_H$
assuming $M_c=1.45$~GeV, $M_b=4.7$~GeV, and $\alpha_S(m_Z)=0.123$.}
\end{table}

The QCD corrections to $\Gamma\left(H\to q\bar q\right)$ are known
up to $O\left(\alpha_S^2\right)$ for $q\ne t$.
In the $\overline{\mbox{MS}}$ scheme, the result is~\cite{Kniehl}
\begin{eqnarray}
\label{hqqmsb}
\Gamma\left(H\to q\bar q\right)&=&{3G_Fm_Hm_q^2\over4\pi\sqrt2}
\left[\left(1-4{m_q^2\over m_H^2}\right)^{3/2}
+{\alpha_S\over\pi}\left({17\over3}-40{m_q^2\over m_H^2}
+O\left({m_q^4\over m_H^4}\right)\right)\right.
\nonumber\\
&&\qquad{}+\left.
\left({\alpha_S\over\pi}\right)^2\left(K_2
+O\left({m_q^2\over m_H^2}\right)\right)
+O\left(\left({\alpha_S\over\pi}\right)^3\right)\right],
\end{eqnarray}
where
 $K_2\approx35.9399-1.3586n_F$~\cite{Gorishny} and it is understood that
$\alpha_S$ and $m_q$ are to be evaluated at $\mu=m_H$.
We note in passing that Eq.~(\ref{hqqmsb}) may be translated into the
on-mass-shell scheme by using the above relation between $M_q$ and
$m_q(m_H)$~\cite{Kat}.
However, appealing to the general notion that the resummation of large
logarithms is automatically implemented by the $\overline{\mbox{MS}}$
evaluation using the appropriate scale,
we express a preference for the use of Eq.~(\ref{hqqmsb}).
The
 difference between these two evaluations is extremely small~\cite{Kat},
which indicates that the residual uncertainty due to the lack of
knowledge of the $O\left(\alpha_S^2m_q^2/m_H^2\right)$ and
$O\left(\alpha_S^3\right)$ terms is likely to be inconsequential for
practical purposes.

The hadronic width of the Higgs boson receives contributions also from
the $H\to gg$ channel, which is mediated by massive-quark triangles,
and related higher-order processes.
The respective partial width is well approximated by~\cite{Djouadi}
  \begin{equation}
  \Gamma\left(H\rightarrow~gg(g),gq\bar q \right)=\Gamma(H\rightarrow~gg)
  \left(1+\frac{\alpha_S(m_H)}{\pi}
  \left(\frac{95}{4}-\frac{7}{6}n_F\right)\right),
\end{equation}
where, in the $m_H$ range of current interest, $n_F=5$ and~\cite{JEllis}
\begin{equation}
\Gamma(H\rightarrow~gg)=\frac{\alpha_S^2(m_H)G_Fm_H^3}
{36\pi^2\sqrt 2}\left(1+\frac{7}{60}\frac{m_H^2}{M_t^2}+
O\left(\frac{m_H^4}{M_t^4}\right)\right).
\end{equation}

Finally, we discuss the one-loop electroweak corrections to the
$H\to f\bar f$ decay rates~\cite{Hff,Bardin}.
In the $m_H$ range under consideration, these may be incorporated by
multiplying Eq.~(\ref{EQ-Gamma}) with~\cite{Hff}
\begin{eqnarray}
{\cal K}&=&\left\{1
+{\alpha\over\pi}\,{3\over2}Q_f^2\left({3\over2}-\ln{m_H^2\over m_f^2}
\right)+{G_F\over8\pi^2\sqrt2}\left[C_fM_t^2
+m_W^2\left(3{\ln\cos^2\theta_W\over\sin^2\theta_W}-5\right)
\right.\right.\nonumber\\&&{}+\left.\left.
\vphantom{\left(3{\ln\cos^2\theta_W\over\sin^2\theta_W}-5\right)}
m_Z^2\left({1\over2}-3\left(1-4\sin^2\theta_W|Q_f|\right)^2\right)\right]
\right\},
\end{eqnarray}
where $Q_f$ is
 the electric charge of $f$ (in units of the positron charge),
$C_b=1$, and $C_f=7$ for $f\ne t,b$.
Note that for $f=b$ the $M_t$ dependence is strongly reduced due to a
cancellation
 between universal self-energy diagrams and specific triangles
involving virtual top quarks.
In the case of $f\ne t,b$, also the two-loop corrections of
$O\left(\alpha_SG_FM_t^2\right)$ are known~\cite{Hll}.
They screen somewhat the one-loop $M_t$ dependence, so that effectively
$C_f=7-2(\pi/3+3/\pi)\alpha_S\approx7-4\alpha_S$.
Electroweak corrections to the fermionic decay rates of the Higgs boson
are relatively modest in the $m_H$ range accessible at LEP200,
the maximum effect being 1.2\% in the $\tau^+\tau^-$ channel,
0.3\% for $c\bar c$, and 0.6\% for $b\bar b$.

We are now in a position to compute accurately all the Higgs-boson
branching ratios that are of phenomenological relevance at LEPI and
LEP200.
Our final
 results are summarized in Fig.~\ref{FIG-2} and Table~\ref{TAB-3}.
\begin{figure}
\caption{Branching fractions of the Higgs boson in the $m_H$ window
relevant for LEPI and LEP200.
All radiative corrections discussed in the text are included.}
\label{FIG-2}
\end{figure}
\begin{table}[tbp]
  \begin{center}
    \begin{tabular}{|c||c|c|c|c|c|c|c||}\hline
$m_H$ [GeV] & $b\bar b$ & $\tau^+\tau^-$ & $c\bar c$ & $gg$ & $W^*W^*$ &
$Z^*Z^*$ & $\gamma\gamma$  \\ \hline\hline
 50 & 88.3 & 8.7 & 2.8 & 0.2 &  -  &  -  &  -   \\ \hline
 60 & 87.9 & 9.0 & 2.8 & 0.3 &  -  &  -  &  -   \\ \hline
 70 & 87.6 & 9.2 & 2.8 & 0.4 &  -  &  -  &  -   \\ \hline
 80 & 87.1 & 9.5 & 2.8 & 0.4 & 0.1 &  -  & 0.1  \\ \hline
 90 & 86.7 & 9.6 & 2.8 & 0.6 & 0.2 &  -  & 0.1  \\ \hline
100 & 85.3 & 9.7 & 2.7 & 0.7 & 1.3 & 0.1 & 0.2  \\ \hline
110 & 80.8 & 9.4 & 2.6 & 0.8 & 5.7 & 0.5 & 0.2  \\ \hline
    \hline\end{tabular}\newline
  \end{center}
  \caption[The Running Quark Masses]
  {\label{TAB-3}
Branching fractions (in \%) of the Higgs boson in the $m_H$ window
relevant for LEPI and LEP200.
All radiative corrections discussed in the text are included.}
\end{table}
To appreciate the effect of the radiative corrections to the decay widths
on the branching fractions,
we contrast in Table~\ref{TAB-2} our final results for $m_H=80$~GeV
with the
 evaluations based on the tree-level decay rates along with the quark
pole masses according to Eq.~(\ref{EQ-BRs}).
The sole effect of running the quark masses in the tree level formulae is
also shown. One can see that the latter makes up the dominant effect on
the resulting branching ratios.
\begin{table}[tbp]
  \begin{center}
    \begin{tabular}{|c||c|c|c||}\hline
 Decay Mode & Tree-Level & $m_q\rightarrow m_q(m_H)$
                                   & Full \\ \hline \hline
 $H\to b\bar b$       &  87.3      &  85.5    &    87.1         \\ \hline
 $H\to\tau^+\tau^-$   &   4.2      &  11.8    &     9.5         \\ \hline
 $H\to c\bar c$       &   8.5      &   2.7    &     2.8         \\ \hline
 $H\to gg$            &    -       &   -      &     0.4         \\ \hline
 $H\to W^*W^*$        &    -       &   -      &     0.1         \\ \hline
 $H\to \gamma\gamma$  &    -       &   -      &     0.1         \\ \hline
    \hline\end{tabular}\newline
  \end{center}
  \caption[$BR(H\rightarrow ALL)$]
  {\label{TAB-2} Branching fractions (in \%) of an 80~GeV Higgs boson
  evaluated
   from Eq.~(\ref{EQ-BRs}) with $M_q$ and $m_q(m_H)$, respectively,
  and evaluated including all channels and radiative corrections
  discussed in the text.}
\end{table}
 \subsection{Systematics}

We do not include the $H\rightarrow s\bar s$ channel in our analysis,
since its partial width is greatly suppressed by the smallness of
$m_s(m_H)$.
However, uncertainties in $M_c$, $M_b$, $M_t$, and $\alpha_S(m_Z)$
are found to be significant.
Their maximum effects on the branching ratios are investigated
in Table~\ref{TAB-6} for $m_H=80$~GeV.
The numbers are very similar for other values of $m_H$.
\begin{table}[h]
  \begin{center}
    \begin{tabular}{|c|c||c|c|c|c||}\hline
Decay Mode& $BR$ [\%]
 & $M_b=(4.7$ & $ M_c=(1.45$ & $\alpha_S(m_Z)=0.123$ & $M_t=(165$ \\
 & & $\pm 0.2)$~GeV & $\pm 0.05)$~GeV & $\pm 0.006$ & $\pm 35)$~GeV
 \\ \hline \hline
$H\rightarrow~b\bar b$      & 87.1 & $^{+1.2}_{-1.3}$ &
 $^{+0.4}_{-0.3} $ & $^{+0.0}_{-0.1} $ & $^{+0.1}_{-0.0}$
  \\ \hline
$H\rightarrow~c\bar c$      & 2.8  & $^{+0.3}_{-0.3}$ &
 $^{+0.4}_{-0.4} $ & $^{+0.9}_{-1.1} $ &$^{+0.0}_{-0.0} $
  \\ \hline
$H\rightarrow~gg$           & 0.4  & $^{+0.1}_{-0.0}$ &
 $^{+0.1}_{-0.1} $ & $^{+0.1}_{-0.0} $ &$^{+0.0}_{-0.0}$
  \\ \hline
$H\rightarrow~\tau^+\tau^-$ & 9.4  & $^{+1.0}_{-0.8}$ &
 $^{+0.1}_{-0.0} $ & $^{+1.0}_{-0.8} $ &$^{+0.1}_{-0.0}$
  \\ \hline
    \hline\end{tabular}\newline
  \end{center}
  \caption[Systematics]
  {\label{TAB-6} Effects of the uncertainties in $M_q$
    and $\alpha_S(m_Z)$ on the various Higgs branching ratios for
    $m_H=80.0~GeV$.}
\end{table}
 The total systematics is obtained by combining the individual errors
 in quadrature.
 The present situation is visualized in Fig.~\ref{FIG-44}.
\begin{figure}
\caption{Branching ratios of the Higgs-boson decays to (a) $b\bar b$,
(b) $\tau^+\tau^-$, $c\bar c$, $W^*W^*$,
(c) $gg$, $\gamma\gamma$, $\gamma Z$, and $Z^*Z^*$
versus $m_H$.
The bands indicate the systematic errors.}
\label{FIG-44}
\end{figure}
 \section{Conclusion}

In this paper, we have collected and updated the theoretical predictions
for the
 Higgs-boson production rate and decay branching fractions appropriate
to LEP200 conditions.
Our analysis
of $\sigma\left(e^+e^-\to Hf\bar f\right)$ includes initial-state
bremsstrahlung to second order with exponentiation, finite-width effects,
and the full one-loop weak corrections to the underlying $2\to2$ process,
$e^+e^-\to HZ$.
We showed that
 the popular Improved Born Approximation supplemented with the
same initial-state corrections deviates appreciably from the evaluation
with the full weak corrections, especially at energies far above the
threshold of on-shell $HZ$ production.
We also suggested
a way to implement weak corrections below this threshold.
We have assigned a 0.9\% systematical error to our results
arising mainly from the uncertainties in $m_Z$, $\Gamma_Z$, and $M_t$.

In our analysis of the Higgs-boson branching fractions, we took into
account two-loop QCD corrections to the hadronic widths,
one-loop electroweak corrections to the fermionic widths,
as well as the
 contributions from the $\gamma\gamma$, $\gamma Z$, $Z^*Z^*$,
and $W^*W^*$ channels.
The branching ratios of $H\rightarrow b\bar b$ and
$H\rightarrow~\tau^+\tau^-$ were found to differ slightly from the
commonly quoted numbers.
{\it E.g.}, for $m_H=80$~GeV, we obtained
$BR\left(H\rightarrow~b\bar b\right)=(87.1\pm 1.3)\%$
and $BR(H\rightarrow\tau^+\tau^-)=(9.4^{+1.4}_{-1.1})\%$.
We also estimated the systematics on the branching fractions,
which should be relevant already for the determination of the
$m_H$ lower bound at LEPI.

\vskip2cm
\noindent
{\bf Acknowledgements.}
We would like to thank Ehud Duchovni, Andrei Kataev, Victor Kim,
Yossi Nir, and Peter Zerwas for useful communications.
E.G. is an incumbent of the Jack and Florence Goodman career development
chair and
is partly supported by the Robert Rees fund for applied research.
B.K. is grateful to the Albert Einstein Center for Theoretical Physics at
the Weizmann Institute of Science for its hospitality
and for supporting his visit,
during which part of this work was carried out.

\end{document}